\documentclass[12pt]{article}
\usepackage{amsmath,amssymb,graphicx,mathrsfs,slashed}
\newcommand{\be}{\begin{equation}}
\newcommand{\ee}{\end{equation}}
\newcommand{\bea}{\begin{eqnarray}}
\newcommand{\eea}{\end{eqnarray}}

\newcommand{\wh}{\widehat}

\def\({\left(} \def\){\right)}

\begin{document}
\title{\vspace{-1.8in}
{Falling through the  black hole horizon}}
\author{\large Ram Brustein${}^{(1)}$,  A.J.M. Medved${}^{(2,3)}$ \\
\vspace{-.5in} \hspace{-1.5in} \vbox{
 \begin{flushleft}
  $^{\textrm{\normalsize
(1)\ Department of Physics, Ben-Gurion University,
    Beer-Sheva 84105, Israel}}$
$^{\textrm{\normalsize (2)  Department of Physics \& Electronics, Rhodes University,
  Grahamstown 6140, South Africa}}$
$^{\textrm{\normalsize (3)
National Institute for Theoretical Physics (NITheP), Western Cape,
South Africa
 }}$
 \\ \small \hspace{1.7in}
    ramyb@bgu.ac.il,\  j.medved@ru.ac.za
\end{flushleft}
}}
\date{}
\maketitle
\begin{abstract}
We consider the fate of a small classical object, a ``stick'', as it falls  through the horizon of a  large black hole (BH). Classically, the equivalence principle dictates that the stick is affected by small tidal forces, and Hawking's quantum-mechanical model of BH evaporation makes essentially the same prediction. If, on the other hand, the BH horizon is surrounded by a ``firewall'', the stick will be consumed as it falls through.  We have recently extended Hawking's model by taking into account the quantum fluctuations of the geometry and the classical back-reaction of the emitted particles. Here, we calculate the strain exerted on the falling stick for our model. The strain depends on the near-horizon state of the Hawking pairs. We find that, after the Page time when the state of the pairs deviates significantly from maximal entanglement (as required by unitarity),  the induced strain in our semiclassical model is still parametrically small. This is because the number of the disentangled pairs is parametrically smaller than the BH entropy. A firewall does, however, appear if the number of disentangled pairs near the horizon is of order of the BH entropy, as implicitly assumed in previous discussions in the literature.
\end{abstract}
\newpage

\section{Introduction}

What would be the fate of a small classical object as it falls through the horizon of a  large black hole (BH)?  Given that the BH is large enough, classical
relativity predicts that  the object  will only suffer a small tidal force.
As  Hawking's quantum-mechanical model of BH evaporation makes essentially the same prediction \cite{Hawk,info}, this was long thought to be a settled matter.
Nonetheless, the  table has since  been turned on account of the recent ``firewall''
proposal, which suggests that the object will rather be obliterated
due to interactions with high-energy quanta \cite{AMPS}.
 Also see \cite{Sunny1,info4,Braun} for earlier, related discussions, \cite{MP,Bousso2,Mathur} for important clarifications and \cite{fw1,fw2,fw3,fw4,Sunny2,avery,lowe,vv,pap,Giddings,jac,HH,Avery2,AMPSS,lowefw2,SMfw,pagefw,MT,VRfw,newfw3,newhawk,Harlow}
for what is just  a sampling of the firewall literature.

We have recently developed a new semiclassical model of BH  evaporation and  would now like to find out what our model predicts for the fate of a falling object.  We have so far studied the model both from the perspective of the emitted Hawking radiation as observed from outside the BH \cite{slowleak,slowburn,endgame,density} and from the  perspective of pair production near the BH horizon \cite{flameoff,noburn,schwing}. The premise has been to extend  Hawking's original models --- respectively, the  collapsing matter shell \cite{Hawk} and the eternal BH \cite{info}  --- in a way that incorporates the quantum fluctuations of the background geometry and the back-reaction effects of the produced pairs.

The main idea of  our semiclassical model is that one has to treat the BH as a quantum state \cite{RB} rather than a fixed classical geometry. This induces corrections that are non-perturbative from the perspective of an effective  theory of quantum fields on a fixed curved background but, yet,  can significantly alter the outcomes. The analysis is carried out by introducing a Gaussian wavefunction for the horizon of the (incipient) BH, as motivated in \cite{RJ,RM,RB,flucyou}, and reevaluating all relevant quantities as expectation values.  In effect, we take into account that the BH is of finite size and monotonically decaying throughout the process.

Here, we are mostly interested in the pair-production point of view. As discussed previously in \cite{flameoff,noburn}, we have realized a picture in which the produced pairs remain in the near-horizon zone a parametrically short time in comparison to the Page time  \cite{page} ({\em i.e.}, the midpoint of evaporation in units of entropy). After this briefer interval of time --- which we have called the coherence time $t_{coh}$ --- the negative-energy modes should be viewed as having been subsumed by the BH interior and their positive-energy partners, as having transitioned to the  external Hawking radiation. Then,  as a consequence of this continual depletion of modes from  the near-horizon zone, the number of pairs in this region is of the order of the square root of the BH entropy.

This last outcome should, in a qualitative sense, really be regarded as generic. After all, given any model for  which the BH mass is finite and  decreasing in time due to the emission of particles, the number of pairs in the near-horizon region should be parametrically smaller than the BH entropy. For instance, by the Page time, about half of the particles that  will ever be emitted by the BH  have  already moved far away from the horizon and transitioned into  ``real'' Hawking particles. Clearly, then, the number of pairs in the zone cannot be any larger than the remaining number of would-be Hawking particles that are still waiting to be emitted.

We will show that this estimate for the number of pairs implies that they induce a parametrically small force on free-falling objects crossing the horizon. This is in direct  contrast to  the  aforementioned firewall proposal.

The basic idea underlying the firewall proposal is as follows: The standard properties of quantum mechanics, such as unitary evolution and the strong subadditivity of  entropy, prohibit the positive-energy modes in the zone from being concurrently entangled with both their negative-energy partners and the older, outgoing Hawking particles. The former is necessary to ensure that the horizon is free of drama, while the latter is needed  for the eventual purification of the radiation. This conclusion is indeed correct and requires the produced pairs in our model to deviate significantly from maximal entanglement at times later than the Page time \cite{schwing}. However, this observation by itself does not determine the number of disentangled pairs near the horizon and, hence, the amount of excitation in the near-horizon region above the Hartle-Hawking vacuum.

We have previously studied the plight of an in-falling shell of  matter (quantum or classical) as it passes through the horizon \cite{noburn}. Our findings revealed that the shell ``sees''~\footnote{More accurately, what an external observer perceives the shell to see.} excitations of the vacuum that are  parametrically suppressed relative  to the Planckian energy scales that are normally attributed to  a  firewall. However, this analysis was limited in  the following three ways: First, our calculation  was based simply on estimating the magnitude of the energy density near the horizon. What really is needed would be a physical result that can be directly compared to the  situation when no excited modes are present. Second, we would like to re-express the situation as much as possible from the direct perspective of the falling object. This is a non-trivial extension because  our framework --- just like Hawking's --- is formulated from the perspective of an external, stationary observer. Third,  we had not yet accounted for the possibility of strong deviations from maximal entanglement and the resulting properties of the state of the matter fields in the near-horizon zone. As it turns out,  we only need some limited information about the near-horizon state and  do not need to know the state of the interior of the BH.

\subsection{A thought experiment and its outcome}

We are proposing a thought experiment that consists of dropping a cylindrical ``stick'' radially towards the BH horizon and asking how its journey  is influenced  by the disentangled modes within the near-horizon region. We can calculate the total number of pairs in this region  and the degree of disentanglement amongst them. These inputs enable us to determine the curvature  that is induced  by the disentangled modes and then, by way of the geodesic deviation equation, the corresponding force on the stick in terms of a dimensionless parameter, the mechanical strain. Then we can discuss whether the induced  strain can be used to detect the position of the horizon and to what extent, if at all, the falling stick is consumed as it falls through this surface.

The current approach allows us to discuss the geodesic deviation equation from the perspective of the falling stick. Hence, there is no longer any need to speculate as to what is the precise definition of the firewall, which remains elusive. We can compare the gravitational force delivered  by the disentangled modes to that delivered by the background Riemann curvature and discuss the implications. This can be done for Hawking's model, for our semiclassical model and for the Page model \cite{page} as implicitly interpreted in the context of the firewall discussions. As for the force induced by additional interactions, such as those of  electromagnetism, it is likely to be subdominant, but this issue should probably
 be considered in more detail.

We find that the exerted force is proportional to the number of disentangled modes in the vicinity of the horizon and to the amount of disentanglement. After the Page time, the amount of disentanglement per mode is of order unity and, consequently, the force is proportional to the number of Hawking pairs in the near-horizon region. We find for our model that the force delivered by the modes is parametrically larger than that of the background. On the other hand, the strain on the stick is still  parametrically  small  --- it is suppressed by the ratio of the length of the stick to the BH radius. This smallness can be attributed, once again, to the bounded number of pairs in the near-horizon region.

In the Page model \cite{page}, as implicitly interpreted in the firewall discussions, the number of  pairs becomes of order of the BH entropy
at the Page time and the degree  of disentanglement per mode grows to
order unity by the same time \cite{MP}.
As a result,  the force on the stick becomes Planckian, inducing a parametrically large strain which is so large that the stick indeed breaks up. So that, in this case, we find a phenomenon whose outcome leads to a disintegration of the stick and could certainly  be interpreted as a ``firewall". In the Hawking model,
on the other hand, the modes are always maximally entangled up to small corrections, and so  their impact on the stick is much smaller than that of
the background curvature.

In summary, the arguments for a firewall in \cite{AMPS}  are basically substantiated by our results, since the near-horizon state has to be different than that which is  predicted by an effective theory of fields on a fixed BH background. But, at least for our semiclassical model, the degree of deviation from the standard vacuum is much smaller than claimed.

\subsection{Comparison with a previous analysis}

The current treatment was  motivated in part by that  of Itzhaki \cite{Sunny1}, which can be viewed as the first realization of what only later was dubbed a BH firewall. Itzhaki posed the following question: What is the effect of a gravitational shock wave due to an outgoing  Hawking mode on an ingoing test particle? This was computed and the answer was summed over all such  shock waves that the test particle encounters on its way to the horizon. Itzhaki found  that the net effect is to displace the particle  so far from its original (null) trajectory that it never even has the opportunity to cross the horizon --- the BH had already evaporated before the particle ever got there.

This is a remarkable finding and obviously a much  different one than ours. But we believe that there is no contradiction. Itzhaki's conclusion is based on the exponential squeezing of the modes in the vicinity of the horizon; in other words, the exponentially large near-horizon redshift in Hawking's model. The result is that, before the test particle ever reaches the horizon, it crosses the path of {\em all} the emitted Hawking particles.

We, however, view this infinite redshift as an approximation of treating the background geometry as a strictly classical entity \cite{flucyou} (which is tantamount to assuming an infinitely massive BH) and the test particle as strictly point like.  In our framework, the  quantum fluctuations of the BH  regulate this would-be infinite redshift. In fact, as will be shown, the redshift is a ``red herring'' --- the piling-up  of modes near the horizon is  mitigated by the continual depletion
of incipient Hawking particles  from the near-horizon zone, insofar as  the redshift has been suitably regulated.

As we will also be discussed later, our proposal for the energy density of the disentangled modes  is parametrically larger than in Itzhaki's model but,
in spite of this difference, the induced gravitational interactions on a finite object are still small.

\subsection*{Contents}

The rest of the paper is organized as follows: The next section contains a brief explanation of our semiclassical model of BH evaporation. Then, in Section~3, we  use a novel physical argument to affirm our previous description of the
pair-produced modes  in \cite{flameoff};  namely, that the would-be Hawking
particles ``escape'' from the near-horizon region after an  interval of order
$t_{coh}$, which is parametrically shorter than the BH lifetime. The quantitative analysis of the induced strain is found  in Section~4, where we give a detailed account of the plight of the stick. Section~5 contains a brief summary.

\section{The semiclassical model}

In the following, fundamental constants, besides the Planck length $\;l_P=\sqrt{\hbar G}\;$,  are usually set to unity except when
needed  for clarity. We are mostly interested in parametric dependence and so typically neglect numerical factors.

We assume, for concreteness, a four-dimensional Schwarzschild BH with metric $\;ds^2=-F(r)dudv+ r^2d\Omega^2\;$, where $\;F(r)=1-R_S/r\;$ and  $\;R_S=2l_P^2M\;$ is the Schwarzschild radius. Also, $u,v$ are the retarded and advanced null coordinates, $\;u,v=t\mp r_{\ast}\;$, such that $\;r_{\ast}=\int \frac{dr}{F(r)}\;$ is the Tortoise coordinate. The BH entropy is $\;S_{BH}=\frac{\pi R_S^2}{l_P^2}\;$ and the BH is semiclassical, $\;S_{BH}\gg 1\;$.

We use $N$  to denote either the cumulative number of particles emitted from the BH or the cumulative number of pairs produced
(these are parametrically the same number) and $N_{pairs}$ to denote the number of pairs in the
near-horizon zone at some
given time.

Our semiclassical model is similar in many respects to the Hawking model of BH evaporation. However, there is a significant difference: The BH is treated as a quantum state and its quantum fluctuations are not neglected. In practice, we achieve this goal by assigning the (incipient) BH a Gaussian wavefunction \cite{RM,RB,flucyou}
\be
\label{solutionfin}
\Psi_{BH}(R)\;=\;{\cal N}^{-1/2} e^{\hbox{$-\frac{\pi}{2 l_P^2} (R-R_S)^2$}}\;,
\ee
where $R$ parametrizes  the fluctuating position of
the quantum horizon and ${\cal N}$ is a normalization constant.
We then calculate quantum expectation values rather than work directly with the classical metric. For an observable $\widehat O$, this means calculating
\be
\langle \Psi_{BH}|  \wh{O}  |\Psi_{BH}\rangle
 \;=\; 4\pi
\int\limits^{\infty}_{0}  dR\; R^2 O(R) \Psi^2_{BH}(R)\;.
\label{expval}
 \ee

The small parameter in our model  is the ``classicality'' parameter, $C_{BH}=1/S_{BH}$. What is essentially the same parameter also appears in \cite{Dvali1,Dvali2}. Technically, it is introduced by the width of $\Psi_{BH}$.
The fact that the classicality parameter does not vanish --- it  is rather small but finite --- can result in modifications to physical quantities.
 The differences are most pronounced for quantities that are either vanishing or divergent in the classical limit $C_{BH}=0$.

In anticipation of the upcoming sections, we list here several relevant results:
\bea
\langle \Psi_{BH} |F^2| \Psi_{BH} \rangle &\equiv &
\lim\limits_{r\to R_S}
\langle \Psi_{BH} |\left(\frac{r-R}{R}\right)^2 | \Psi_{BH} \rangle \;\simeq\;
S^{-1}_{BH} \;,
\label{2F} \\
\langle \Psi_{BH} |F^{-2}| \Psi_{BH} \rangle &\equiv &
\lim\limits_{r\to R_S}
\langle \Psi_{BH} |\left(\frac{R}{r-R}\right)^2| \Psi_{BH} \rangle \;\simeq\; S_{BH}\;, \label{F2} \\
\langle \Psi_{BH} |F^{-4}| \Psi_{BH} \rangle &\equiv &
\lim\limits_{r\to R_S}
\langle \Psi_{BH} |\left(\frac{R}{r-R}\right)^4| \Psi_{BH} \rangle
\;\simeq\; S_{BH}^2\;,
\label{4F}
\eea
where the latter two follow from the use of
\be
\int dx \frac{1}{x^{2n}} e^{-S_{BH} x^2} \;\simeq\; \Gamma\left(\frac{-n+1}{2}\right) S_{BH}^n\;.
\ee

An important time scale that accounts for the difference between an infinitely massive BH and one with a finite but large mass is the so-called coherence time. This time scale is typically $\;N_{coh}\simeq\sqrt{S_{BH}}\;$  (which reads in Schwarzschild units as $\; t_{coh}\simeq \frac{R^2_S}{l_P}$) and has a dual meaning: First, from the perspective of the external radiation, $N_{coh}$ is the scale of temporal extent for the matter correlations. This  comes about because  wavefunctions for the BH  at  different times  are  orthogonal when this time difference
exceeds the coherence scale. Second,  from the perspective of the  produced pairs, $N_{coh}$ is the time that a mode stays in the near-horizon zone; after which, the negative-energy modes are subsumed by the interior matter and the positive-energy modes escape to become Hawking particles. It is latter meaning that is significant to the current work and will be substantiated in the section to follow.

This is the bare necessity that a reader needs to know about  our model for BH evaporation.
Our earlier, cited papers can be consulted for more comprehensive discussions.

\section{The Hawking modes near the horizon}

Here, we will reconsider the pair-production  picture of BH evaporation. Many of the aspects are common to the Hawking  and  semiclassical models. We will emphasize these aspects as well as the differences  as they turn up in the discussion.

If quantities are averaged over sufficiently long time periods,  it should be clear that the average number of produced pairs   must match the average number of emitted particles. On average, a Hawking particle is emitted once every Schwarzschild time $\;t\sim R_S\;$,  and so pairs are produced at the same rate. Then, since $\;R_S\ll t_{coh}=R_S \frac{R_S}{l_P}\;$, we can treat both processes as acting continuously when looking at intervals of  coherence time.

In this way, the process of BH evaporation entails the continuous  production of pairs and the continuous absorption  of negative-energy modes. Meanwhile,  positive-energy modes are transitioning into the outgoing Hawking radiation as  their subsumed negative-energy partners are being absorbed into the interior matter, continually reducing the BH mass. All  of these rates are determined, on average, by the BH's thermal rate of emission.

Let us begin the analysis  from the  perspective of a local, free-falling observer. We are interested only in the  massless modes with low angular momenta, which eventually do  escape from the near-horizon zone. These are the modes that are constrained by the arguments of strong subadditivity while the rest  are in their vacuum states.  Each of the massless modes will have a momentum of magnitude $E$, where $\;E\sim 1/R_S\;$, as $R_S$ sets the size of the wavelength. It
then follows, from momentum conservation and from Hawking's realization that the positive- and negative-energy  partners are created (respectively) just outside and just inside of the horizon, that their momenta are initially of the form
\be
\vec{p}\;=\;  \cos{\alpha} E \hat{Y} +   \sin{\alpha} E \hat{U}\;,
\ee
\be
\vec{q}\;=\; - \cos{\alpha} E \hat{Y}-  \sin {\alpha} E \hat{U}\;,
\ee
for  the  positive- and negative-energy mode respectively. Here,
 $\hat {Y}$ defines
a direction along or on the horizon surface (it could be lightlike or spacelike), $\hat {U}$ is the  lightlike Kruskal direction
off the horizon and we are using the conventions that  $U$ increases towards large values of $r$
and $\;0\leq \alpha \leq \pi\;$.

The unit vector $\hat {Y}$ is a linear combination of the unit vectors  $\hat{\theta}$, $\hat{\phi}$  and the null Kruskal direction
along the horizon $\hat {V}$. The exact form of this linear combination is not relevant to
the current considerations. In most cases, $\;\sin{\alpha}$, $|\cos{\alpha}|$  are of order unity and will be dropped for now on.

The above relations are from a local, free-falling perspective.
A stationary observer at large $r$ still detects modes with energy $\;E\sim 1/R_S\;$ but sees the momentum in the $U$ direction
as being red-shifted according to
\be
\vec{p}\;=\;  E\hat{Y} +   e^{-u/2R_S} E \hat{U}\;,
\ee
\be
\vec{q}\;=\; - E \hat{Y} -  e^{-u/2R_S} E  \hat{U}\;,
\ee
where $u$ is the retarded time coordinate and the redshift factors
are meant to account for both the energy and the velocity of the mode.

Now, given the standard classical geometry of Hawking's model, $\;u\simeq -2R_S\ln{\left(\frac{r-R_S}{R_S}\right)}\;$~\footnote{The factor of 2 is because $\;t\sim -r_{\ast}\;$ at the future horizon.}
and then $\;e^{-u/2R_S}\simeq \frac{r-R_S}{R_S}\to 0\;$ as
$\;r\to R_S\;$. This makes it clear that, from  a stationary-observer's
viewpoint,  the $U$ component of the momentum $\vec{P}_{U}$ vanishes and so  the partners are forever trapped on the horizon. This is consistent with Hawking's description of the pair-production process  in \cite{info}, as
there an eternal BH spacetime is assumed.

But our semiclassical model leads to a different result. The average $\langle \Psi_{BH}|  \vec p_U   |\Psi_{BH}\rangle$ is still exponentially small; however the
quantum fluctuations of the BH itself will lead to a small but finite variance \cite{flucyou}.
Indeed, using the prescription~(\ref{expval}) and the result~(\ref{2F}),
we find that
\bea
\langle \Psi_{BH}|  \vec p_U\cdot\vec p_U   |\Psi_{BH}\rangle &=& E^2
\langle \Psi_{BH} |F^2| \Psi_{BH} \rangle  \nonumber \\
&=& \frac{E^2}{S_{BH}} +{\cal O}[S_{BH}^{-2}]\;.
\label{approx}
\eea

Now, since the average value of $\vec p_U$ is exponentially small and these modes are outgoing so that $\vec p_U$  positive by definition, we can use  $\;\sqrt{\langle | \vec p_U\cdot\vec p_U |\rangle}\simeq E\frac{l_p}{R_S}\;$ as an estimate for the velocity
of a mode in  the direction orthogonal to the horizon,
\be
v_{U}\;=\;\frac{\sqrt{\langle |  \vec p_U\cdot\vec p_U  |\rangle}}{E}\;\simeq\; \frac{l_p}{R_S} \;.
\ee
We can then quantify the time of escape by using the above estimate.
It follows that the time a Hawking mode takes to reach a distance $R_S$ away from the horizon is given by
\be
t_{\rm escape}\;\simeq\; \frac{R_S}{v_U}\;\simeq\; R_S \frac{R_S}{l_P}\;=\;t_{coh}\;.
\ee

In effect, the normally divergent factor in the escape-time estimate has been replaced by the large but finite factor $R_S/l_P$. To summarize, the effect of the quantum fluctuations is to make the redshift finite --- it takes the outgoing mode a finite time to escape the near-horizon region.

The  number of actively entangled pairs $N_{pairs}$ is then of the same order
as the number of pairs that are  produced by the BH over a time period $t_{coh}$; {\em i.e.},  $\;N_{pairs}\sim N_{coh}\;$. As commented upon earlier and detailed elsewhere \cite{flameoff,noburn}, this
truncation in the number of  partnered modes ---  from order  $S_{BH}$ to order $\;N_{coh}=S^{1/2}_{BH}\;$ --- is a critical part of our argument for resolving the aforementioned firewall problem.

Another important quantity that we would like to introduce is
$N_{dis}$, the  number of active pairs times the degree of disentanglement
 per pair ${\cal D}_{dis}$.
This is a model-dependent outcome,  as it requires
a specification of the state of the pairs or, at the very least,
some means of quantifying how much this state deviates from
the Hartle--Hawking state  of maximal entanglement.
Let us first recall that, for our model,  $N_{pairs}$ is  bounded from above
by $\;N_{coh}=\sqrt{S_{BH}}\;$ and so $\;N_{dis}\lesssim \sqrt{S_{BH}}\;$.

In addition, we have recently \cite{schwing} found a means for  estimating ${\cal D}_{dis}$  in our framework. By partitioning the system of Hawking modes into three subsystems --- the already emitted Hawking particles (or early radiation) $A$, the  positive-energy modes in the zone (or late radiation) $B$  and their negative-energy partners $C$ --- we have evaluated the entanglement between $A$ and $B$~\footnote{We used Renyi entropies for the analysis, with this choice justified in \cite{schwing}.}. The condition of strong subadditivity of entropy then  enforces a lower bound on the  the degree of disentanglement between a pair of modes in  $B$ and $C$. The need for such  a bound is quite  natural, given that  ``monogamy of entanglement'' is in play and that any  positive-energy mode in the zone must have some degree of  entanglement with subset $A$ if the state of the radiation is to eventually purify.  What the analysis in \cite{schwing} does is put this idea on a more quantitative level.

And so, taking this  lower bound as an estimate for
the degree of disentanglement per pair, we  have
\be
{\cal D}_{dis}\; \simeq \;\frac{ N C_{BH}-1}{1+ N C_{BH}}\;.
\label{Fdis}
\ee
One can notice  that  ${\cal D}_{dis}$  depends solely  on
the product $NC_{BH}=N(t)/S_{BH}(t)$, which happens to be the effective perturbative parameter
for  our framework \cite{slowleak}. One can also see that,  after the Page time when $\;NC_{BH}\geq 1\;$, the amount of disentanglement is of order unity, $\;{\cal D}_{dis}\sim 1\;$.

As a lower bound, this estimate is not useful at times
before the Page time, for which ${\cal D}_{dis}$ is negative.
However, when the BH is still young,
 ${\cal D}_{dis}$ must be a parametrically small number, as
 any model of BH evaporation should reduce to  Hawking's model plus perturbatively small corrections at such early times. We will always  be assuming
that the BH is older than the Page time, as this is the regime of
interest as far as the prospects for a firewall are concerned \cite{AMPS}.

\section{The fate of a falling stick}

\subsection{Setup of the thought experiment}

We will next consider  the consequences of our framework for a stick falling through the horizon  of a
semiclassical BH.
By  stick, we mean  a classical, cylindrical object  of length $\ell$ and radius $s$, such that $\;l_P\ll s \lesssim \ell \ll R_S\;$. The local
frame of the stick  will  be denoted by $T$, $X$, $Y$ and $Z$.
It is assumed to be  falling toward the horizon on a radial trajectory with its  long side aligned parallel to the direction of motion --- decreasing $r$ or, locally for the stick, decreasing $X$. In Fig.~1, the falling stick is depicted in a space-time diagram.

Before proceeding, we need to make sure that the stick does not disintegrate due to any tidal forces arising from  the   gravitational background. This requirement imposes a constraint on the ratio $\ell/R_S\;$; {\em i.e.}, the ratio of the size of the stick  to the Schwarzschild radius  of the BH.
This comes about because the stick will experience a relative longitudinal acceleration between its ends, for which the magnitude  near the horizon is given by
$\;\Delta a \sim \frac{GM}{R_S^2}\frac{\ell}{R_S}\;$.

To continue with this
idea, let us assume  that the stick is made of some elastic material; then
$\;\Delta a \sim \frac{K}{\rho} \Delta \ell\;$, where $\rho$ is the mass density of the stick and $K$ is its bulk modulus.
It follows that, near the horizon, $\;\frac{K}{\rho} \Delta \ell \sim
\frac{GM}{R_S^2} ({\ell}/ {R_S})\;$  or $\;{\Delta \ell}/ {\ell} \sim
({\rho}/ {K}) ({c^2}/ {R_S^2})\;$. Here, we have reinstated the speed of light $c$ and used that $R_S=2GM/c^2$.
But $\;({K}/ {\rho}) \sim \omega_{\rm stick}^2\sim ({c_{\rm sound}}/ {\ell})^2 \;$,
and so the result is $\;{\Delta \ell}/ {\ell} \sim ({c}/ {c_{\rm sound}})^2
({\ell}/ {R_S})^2\;$.  Meaning that, if we insist upon $\;{\Delta \ell}/ {\ell} < 1\;$,  then $\;{\ell}/ {R_S} < {c_{\rm sound}}/ {c}\;$.
For known materials, this ratio is no larger than about $10^{-4}$. We can then
conclude that,
to avoid the breaking up of the stick, the ratio ${\ell}/ {R_S}$ has to be parametrically small.

\begin{figure}
[t]
\scalebox{.75} {\includegraphics{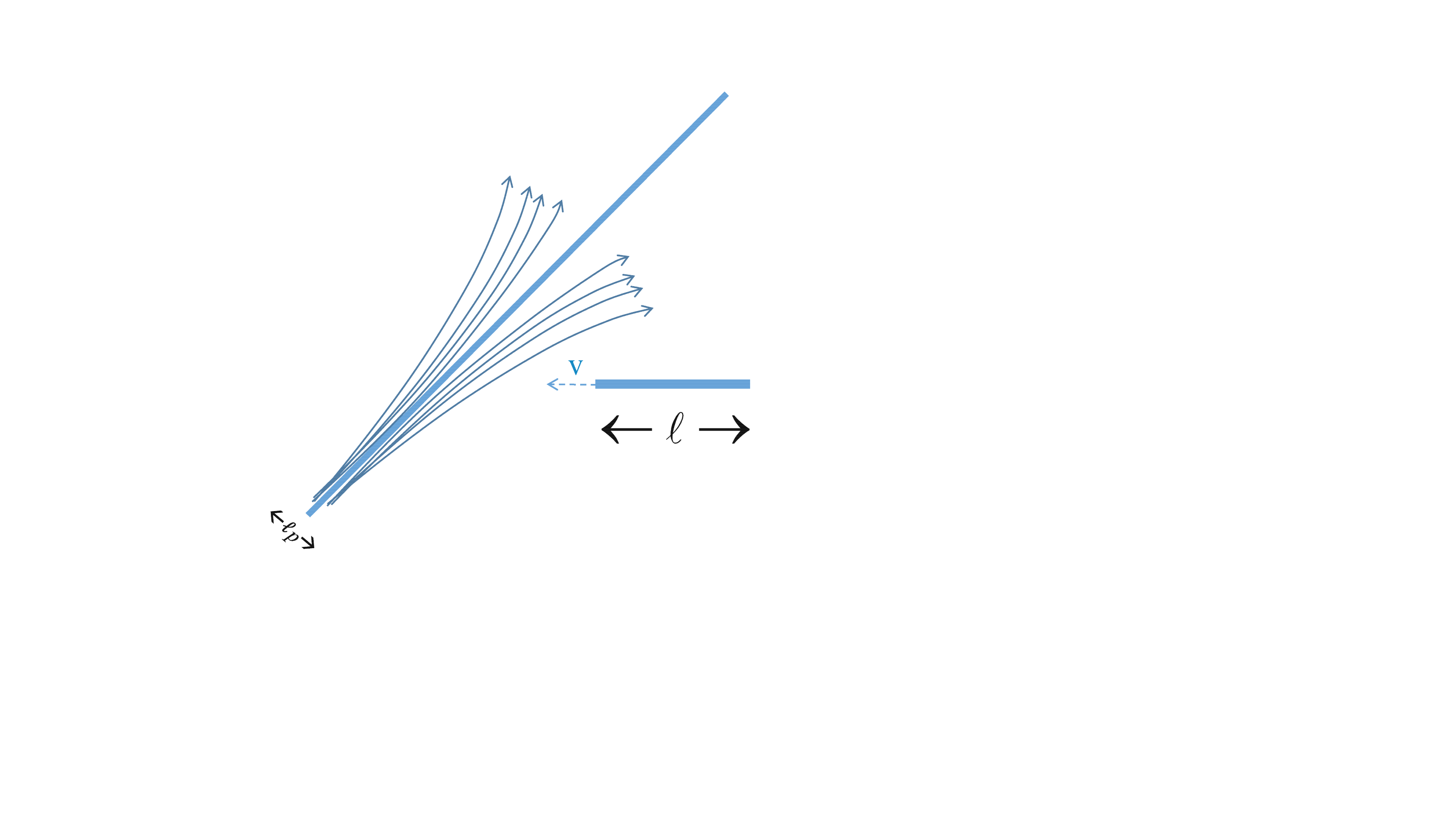}}
\caption{Spacetime diagram showing the Hawking pairs near the BH horizon with the falling stick away from the horizon.}
\end{figure}
\begin{figure}
[t]
\scalebox{0.7} {\includegraphics{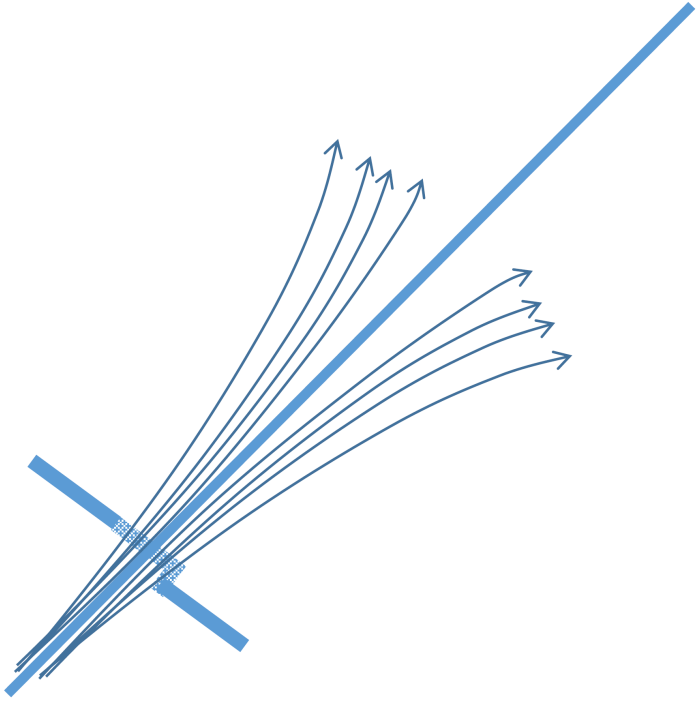}}
\caption{Spacetime diagram showing  the deformation of the stick induced by the disentangled Hawking modes as it crosses  the horizon.}
\end{figure}
\begin{figure}
[t]
\scalebox{0.8} {\includegraphics{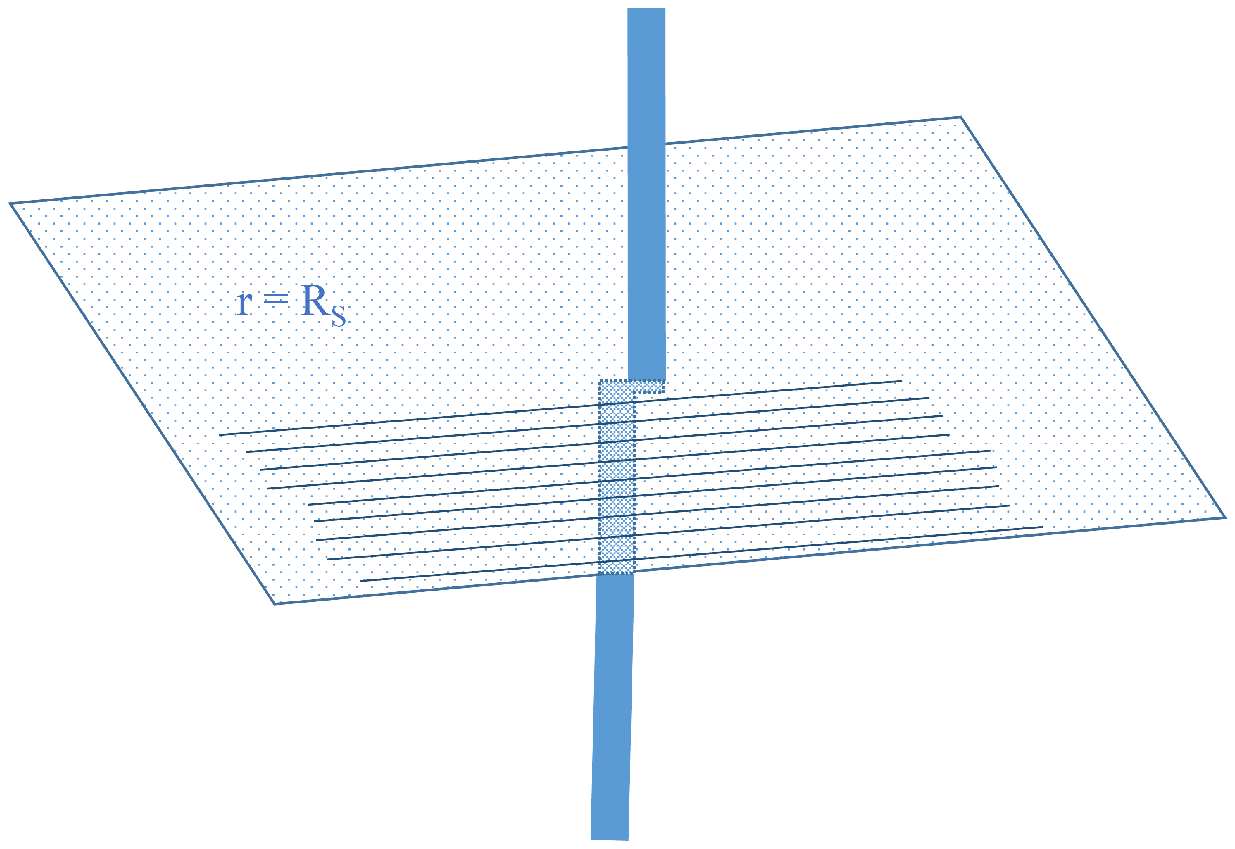}}
\caption{The deformation of the stick induced by the disentangled Hawking modes across the surface $r=R_S$.}
\end{figure}

For this setup, the acceleration  of the stick in the $Y$ (or $Z$) direction can be determined from the geodesic deviation equation for Einstein's gravity. What we want to know, in particular, is the  induced gravitational force which is delivered to the stick by the disentangled Hawking modes near the horizon. This force will eventually be parametrized in terms of a dimensionless scalar quantity, the deformation per unit length or the strain $\gamma$.

Let us pause to comment briefly  on the physical picture. Far away from the horizon --- where the Hawking modes are dilute and their induced  force is weak ---
the net force  exerted on the stick will be negligible. Near the horizon, the situation is different.
The effect of  induced gravity is still relatively weak, but the part of the stick that is closer to the horizon will
feel a stronger force than that which is farther. This is because the Hawking modes become  both denser in number and more energetic  as the horizon is approached. It is this gradient and the accompanying tidal force that could cause the stick  to endure harm.

What we first need to know is the stress--energy tensor for the disentangled modes, from  which Einstein's equation will give us the associated curvature and then the geodesic deviation equation will yield the tidal force. Alternatively, one can deduce the induced change to the metric by treating the Hawking modes as shock waves \cite{Sunny1}.

\subsection{The stress-energy tensor of the disentangled modes}

Knowledge of the stress tensor requires one to know about the number of disentangled modes $N_{dis}$ and the energy density of each of these. A detailed discussion about $N_{dis}$  will be deferred until later. We will determine the stress--energy tensor with respect to an external (stationary) observer's perspective, but the associated Riemann tensor will  be extracted from  a manifestly scalar quantity. In this way,  the remainder  of the calculation can proceed from the stick's own point of view.

A simple dimensional analysis suggests that the energy density of a fully disentangled mode  is  \be
\varepsilon \;\simeq \;\frac{1}{R_S^4 F}\;.
\ee
This expression formally diverges near the horizon; however, the fluctuations of the background regulate the divergence in a similar manner to the way in which  the stretched horizon does. The reasoning for the estimate for $\varepsilon$ is as follows: According to an external  observer, the modes are, up to red-shifting effects, delocalized over a spherical shell of radius $R_S$ and width  $R_S$, while the   energy scale of any given mode is set by the Hawking temperature, $1/R_S$. Hence, the energy density per mode when the redshift is disregarded is $1/R^4_S$. The Tolman redshift  introduces a factor of $F(r)$ into the denominator, as both the inverse of the energy scale and the width of the shell are suppressed by a factor of $\sqrt{F(r)}$.

It then follows that
\be
\delta T_{uu}\;\sim \;\delta T_{vv}\;\sim\; \frac{N_{dis}}{R_S^4 F}\;
\label{stress}
\ee
describes the energy flux for the disentangled modes as far
as an external observer is concerned.

We also need to know the background
stress tensor as would be measured  by  the same external observer and,
for this, employ  the standard Schwarzschild result from \cite{Davies}.
Then, {\em very  near} to the horizon, $\;T_{uu}\sim T_{uv}\sim 0\;$
whereas $T_{vv}\sim -1/R^4_S\;$. As each of these is parametrically much
smaller than the estimate in Eq.~(\ref{stress}) for any $\;N_{dis}>l_P/R_S\;$,
the background tensor can be  disregarded in the subsequent
analysis.

The estimate in Eq.~(\ref{stress}) can be compared to that of Itzhaki \cite{Sunny1}. Working with Kruskal coordinates (for which $\;U\sim -R_S e^{-u/2R_S}\;$ and
$\;V\sim R_Se^{v/2R_S}$),  he proposed that $\;T_{UU}\sim \frac{l_p^2}{U^2}\;$ and $\;T_{VV}\sim 0\;$ near the horizon. But, recalling that
$\;T_{uu}\sim U^2 T_{UU}\;$ and $\;T_{VV} \sim T_{vv}\;$, we see that our proposal for $T_{uu}$ and $T_{vv}$ are both a factor of $1/F$ more divergent at the horizon than their counterparts in \cite{Sunny1}, as well as  those of  the standard
Hartle--Hawking  and Unruh states. This can be attributed to the disentangled modes for our picture being highly concentrated in the proximity of the horizon, as
per  the previous section. From our point of the view,
the modes further removed from the horizon have already ``escaped''.

\subsection{The curvature induced by the disentangled modes}

\subsubsection{Curvature from the stress-energy tensor}

We next want to convert Equation~(\ref{stress}) into a statement about curvature and, as already stated,
work with a scalar quantity. The simplest choice of scalar is
\be
G^{ab} G_{ab} \;=\; l_P^4  T^{ab} T_{ab} \;\simeq\; l_P^4
g^{uv}g^{uv}\delta T_{vv}\delta T_{uu}\;
\label{above}
\ee
where $G_{ab}$ is the Einstein tensor. The  equality on the left follows from  Einstein's equation and the relation on the right follows from the disentangled modes being the dominant source.

In our semiclassical  framework, the expression on the far right should be regarded as an expectation value
with respect to the BH wavefunction; this being the context in which the disentangled modes are revealed.  What we are  then calculating is
the expectation value  of the scalar  $G^{ab}G_{ab}$ with respect to
the same wavefunction, and so it is more appropriate to write
\bea
\langle\Psi_{BH}| G^{ab} G_{ab}|\Psi_{BH}\rangle &=& l_P^4\langle\Psi_{BH}|
g^{uv}g^{uv}\delta T_{vv}\delta T_{uu}|\Psi_{BH}\rangle
\nonumber \\ &\simeq& \;N_{dis}^2\frac{l_P^4}{R_S^8} \langle\Psi_{BH}|
\frac{1}{F^4}|\Psi_{BH}\rangle \nonumber \\
&\simeq & \frac{N^2_{dis}}{R_S^4}\;,
\label{Riemsq}
\eea
where the middle line  follows from Eq.~(\ref{stress}) and  the last line follows
from taking the near-horizon limit along with  Eq.~(\ref{4F}).

One can now get a first hint about the fate of the falling stick by looking at various possibilities for $N_{dis}$. Clearly, the largest possible value for $N_{dis}$ is $S_{BH}$. This assumes that a finite fraction of all the modes that were ever emitted by the BH remain in the vicinity of the horizon. In this case, we find from Eq.~(\ref{Riemsq}) that $\;\langle G^{ab} G_{ab} \rangle_{\Psi_{BH}} \sim \frac{1}{l_P^4}\;$, and so it is likely that the stick disintegrates before it reaches the horizon. But, for our semiclassical model and for an old-enough BH,
 $\;N_{dis}\sim N_{pairs} \sim \sqrt{S_{BH}}\;$, for which $\;\langle G^{ab} G_{ab} \rangle_{\Psi_{BH}}  \sim \frac{1}{R_S^2 l_P^2}\;$. Of course, if $N_{dis}$ is of order unity, then $\;\langle G^{ab} G_{ab} \rangle_{\Psi_{BH}}  \sim \frac{1}{R_S^4}\;$, which cannot be distinguished from the background curvature.

Let us next observe that, because of the Ricci scalar contribution in the Einstein tensor
$\;G^{a}_{\;\;b}=R^{a}_{\;\;b}-\frac{1}{2}Rg^{a}_{\;\;b}\;$, all diagonal components of
the (induced) Riemann tensor will be of roughly the same magnitude;
meaning that the diagonal components of $R^{a}_{\;\;b}$
will scale with $N_{dis}/R_S^2$.
It can then be deduced that, in the stick's own frame where the metric
is regular, the root-mean-square (RMS) value of the Ricci curvature is given by
\be
R^{A}_{\;\;\;B} \;\simeq\; \frac{N_{dis}}{R_S^2}\delta^{A}_{\;\;B}\;,
\label{Riem}
\ee
for $\;A,B=\left\{T,X,Y,Z\right\}\;$.
One can immediately see  that $R^{A}_{\;\;\;B}$ vanishes for the Hawking model (up to the implied background contribution)
since   $\;N_{dis}=0\;$ must be true in this case.

\subsubsection{Curvature from the shock wave approximation}

There is another way to quantify the effect of the disentangled modes acting on the stick.
This would be, following Itzhaki \cite{Sunny1}, to treat the modes as shock waves and estimate the change in the metric and  the curvature due to the  waves crossing the stick.
It is appropriate to use Kruskal coordinates for this calculation if it is to be  from the stick's own perspective.
A shock wave of energy $\;E\sim 1/R_S\;$ propagating outwards along the ray $U=U_0$  will change $g_{UU}$ by an amount $\;\delta g_{UU}\sim G E \delta (U-U_0)\sim l_P^2 /R_S\;\delta(U-U_0)\;$. The large Tolman blueshift for the modes in  our model
can be incorporated by the estimate $\;\delta(U-U_0)\sim 1/U\;$, with
$U$ meant to be within a few Planck lengths from the horizon where $\;U\sim 0\;$.
Then the total displacement  will go as $\;\delta g_{UU}\sim N_{dis} l_P^2/R_S U\;$ (and a similar contribution to $\delta g_{VV}$ for the inward-moving  modes). It can be verified that the deformed metric induces a near-horizon curvature of  order
\be
R^{V}_{\;\;U}\;\sim\; -\frac{1}{R_S}\frac{\partial \delta g_{UU}}{\partial U} \;\sim\; N_{dis} \frac{l_P^2}{R_S^2}
\frac{1}{U^2}\;,
\ee
and similarly for $R^{U}_{\;\;V}$.
Then, since  $\;U\sim R_S F\;$, we may use Eq.~(\ref{F2}) to conclude that the RMS value of the Ricci tensor is given by
\be
R^{V}_{\;\;U}\;\sim\;  \frac{N_{dis}}{R_S^2}\;,
\ee
which is in perfect agreement with the estimate from  Eq.~(\ref{Riem}).

\subsection{The induced strain}

Let us next recall the geodesic deviation equation and apply it to the current
setup,
\be
\frac{d^2(\Delta x)^{A}}{d\tau^2}\;=\;  R^{AB}_{\;\;\;\;\;CD}V_B V^C (\Delta x)^D \;,
\ee
where $({\Delta x})^{D}$ describes the spatial extent of stick --- so that $\;(\Delta x)^{X}=\ell\;$,
 $\;(\Delta x)^{Y}=(\Delta x)^{Z}= s\;$ ---
and
$V^A$  is  the velocity vector for the stick in terms of proper time $\tau$.

Using that the velocity vector for the stick is
$\;V^{A}=-\beta\delta^{A}_{\;\;X}\;$
 for some $\;\beta < 1\;$,
we have
\bea
\frac{d^2(\Delta x)^{A}}{d\tau^2} &=& \beta^2  R^{AX}_{\;\;\;\;\;XD} (\Delta x)^D
\nonumber \\ &\simeq& \beta^2 \left.  R^{A}_{\;\;\;D} (\Delta x)^D \right|_{A,D\neq X}\;,
\eea
where the second line follows from $R^{AX}_{\;\;\;\;\;BX}$ being  the same order as $R^{A}_{\;\;\;B}$.

Next, substituting Eq.~(\ref{Riem}) for the Ricci tensor, we obtain
\be
\frac{d^2(\Delta x)^{A}}{d\tau^2}\; \simeq\; \beta^2 \left.\frac{N_{dis}}{R_S^2}(\Delta x)^A\right|_{A\neq X}\;
\ee
or, after integrating twice,
\be
\frac{\delta (\Delta Y)} {\Delta Y}
\; \simeq\;(\Delta\tau)^2 \beta^2 \frac{N_{dis}}{R_S^2}\;
\label{geo}
\ee
and similarly for $Z$.
Here, $\delta(\Delta Y)$ means the RMS  deformation of the stick in the $Y$ direction, so that the magnitude of the left-hand side is the strain  $\gamma$.

At any given time, only a small (about Planck-sized) segment of the stick is exposed
to the potentially dangerous near-horizon modes. For this reason,
it is appropriate to start with  the force acting on a segment
of length $\;\Delta X\sim \Delta\;$ with $\Delta \gtrsim l_P$. The induced deformation on the stick is depicted in a spacetime diagram in Fig.~2 and at a fixed time in Fig.~3.

The proper time that it takes this segment to
pass through the near-horizon zone is then $\;\Delta \tau \sim  \Delta/\beta\;$.
Given these inputs, Eq.~(\ref{geo}) translates into
\be
\gamma_{\Delta} \;\simeq\;  N_{dis}\frac{\Delta^2}{R_S^2}\;.
\ee

To estimate   the total strain endured by the stick, we will assume the ``worst-case scenario'', in which the individual deformations add coherently.
Then the total strain is simply $\;\ell/\Delta\gtrsim \ell/l_P\;$ times the previous result,
\be
\gamma_{stick} \;\lesssim\;  N_{dis}\frac{l_P\ell}{R_S^2}\;.
\ee
The actual strain will depend on the whether or not the stick oscillates,
its speed of sound and so forth. For instance, a more realistic estimate might rather be to add in quadrature the strains on each part of the stick. Then the result would be
the RMS value $\;\gamma_{stick} \sim \sqrt{\frac{\ell}{\Delta}\left(\gamma_{\Delta}\right)^2}\;$,
but this (or any other) modification  would  only weaken the previous estimate.

\subsection{The induced strain in different models}

Let us start with the case which is implicitly  based on  our previous attempt \cite{noburn} at quantifying the effects of  firewall. There, we incorrectly estimated the disentanglement per mode ${\cal D}_{dis}$ as being equal to the product $\;N_{coh}C_{BH}\ll 1\;$. Then $\;N_{dis}<  1\;$ and
the strain on the stick is
\be
\gamma_{stick} \;\simeq \frac {\ell l_P}{R_S^2}\ll 1\;\;\;\;\;\;[{ N_{dis}<  1}]\;,
\ee
which is vanishing in the classical limit.
This outcome explains the underestimation in our previous study.

Let us now consider what happens  in our  semiclassical model
when the more  accurate estimate of   ${\cal D}_{dis}$ in Eq.~(\ref{Fdis}) is utilized. Then, after the Page time,  $\;{\cal D}_{dis}\simeq 1\;$ and so
$\;N_{dis}\sim N_{pairs}\sim\sqrt{S_{BH}}\sim \frac{R_S}{l_P}\;$.
It follows that
\be
\gamma_{stick} \;\simeq \frac {\ell}{R_S} \ll 1\;\;\;\;\;\;[ N_{dis}\sim  \sqrt{S_{BH}}]\;.
\label{track}
\ee

This is clearly a small number, but how small?
As we have seen before, if  this  experiment is to be conducted in a reasonable way, then this
ratio is constrained by $\;{\ell}/ {R_S}<{c_{\rm sound}}/ {c}\lesssim 10^{-4}\;$. So that, in this case, despite the fact that the energy density is parametrically larger than $1/R_S^4$, the physical effect on the stick is still remarkably small. In other words, the equivalence principle is preserved.

Finally, let us  consider  the Page model, as interpreted in the context of the firewall problem and, in particular, after the Page time.  The
usual interpretation of the  Page model is that  the number of pairs near the horizon is limited only by the original BH entropy and  each of
these has order one disentanglement \cite{MP}. Hence, $\;N_{dis} \sim S_{BH}\;$ and one then obtains
\be
\gamma_{stick} \;\simeq \frac {\ell}{l_P}\gg 1\;\;\;\;\;\;[{ N_{dis}\sim  S_{BH}}]\;.
\ee
Such a  large strain indicates that the stick is obliterated
on its journey through the near-horizon region.
This outcome  can  best  be viewed as further evidence that, given  the assumptions of \cite{AMPS},
a firewall is indeed an inevitable consequence.

\section{Summary}

Using a simple thought experiment, we have investigated how the fate of an in-falling classical object passing through the horizon depends on the state of the near-horizon Hawking radiation.  We verified that our semiclassical   framework for BH evaporation and pair production does not lead to a conflict with the equivalence principle of general relativity (while being consistent with standard quantum theory \cite{schwing}.) In particular, it was shown that, as long as  the experiment of dropping an object through the near-horizon region can be safely carried out, the disentangled Hawking modes  will do nothing further to jeopardize the serenity of the journey. This is true in spite of the disentanglement per mode being of order unity, as required for information to escape from the BH, and can be attributed to  the  Hawking pairs having an effective lifetime that is parametrically smaller than the Page time. On the other hand, the Page model, as normally interpreted in the firewall literature, does lead to a conflict with the equivalence principle, thus substantiating the arguments of   \cite{AMPS} and others.

\section*{Acknowledgments}

We thank Sunny Itzhaki for many useful discussions and insights. The research of RB was supported by the Israel Science Foundation grant no. 239/10.
 The research of AJMM received support from an NRF Incentive Funding Grant 85353, an NRF Competitive
Programme Grant 93595 and Rhodes Research Discretionary Grants. AJMM thanks Ben Gurion University for their hospitality during his visit.

\end{document}